\documentclass[aps,preprint,showpacs,preprintnumbers,showkeys,eqsecnum,amsmath,amssymb]{revtex4}

\usepackage{graphics}
\usepackage{graphicx}
\usepackage{txfonts}
\usepackage{dcolumn}
\usepackage{mathrsfs}% Align table columns on decimal point
\usepackage{bm}% bold math
\usepackage{amsmath,amssymb,epsfig,float}

\newcommand{\bq}{\begin{equation}}
\newcommand{\eq}{\end{equation}}
\newcommand{\bqn}{\begin{eqnarray}}
\newcommand{\eqn}{\end{eqnarray}}

\newcommand{\nb}{\nonumber}

\begin{document}

\title{Condensation for non-relativistic matter in Ho\v{r}ava-Lifshitz gravity}

\author{ Jiliang {Jing}\footnote{Electronic address:
jljing@hunnu.edu.cn}}
\author{Songbai Chen}
\author{Qiyuan Pan}

\affiliation{Department of Physics, and Key Laboratory of Low Dimensional Quantum Structures and
Quantum Control of Ministry of Education, Hunan Normal University,
Changsha, Hunan 410081, P. R. China}

\vspace*{0.2cm}
\begin{abstract}

We study condensation for non-relativistic matter in a Ho\v{r}ava-Lifshitz black hole without the condition of the detailed balance. We show that, for the fixed non-relativistic parameter $\alpha_2$ (or the detailed balance parameter $\epsilon$), it is easier for the scalar hair to form as the parameter $\epsilon$ (or $\alpha_2$) becomes larger, but the condensation is not affected by the non-relativistic parameter $\beta_2$. We also find that the ratio of the gap frequency in conductivity to the critical temperature
decreases with the increase of $\epsilon$ and $\alpha_2$, but increases with the increase of $\beta_2$.
The ratio can reduce to the Horowitz-Roberts relation $\omega_g/T_c\approx 8$  obtained in the Einstein gravity and Cai's result $\omega_g/T_c\approx 13$ found in a Ho\v{r}ava-Lifshitz gravity with the condition of the detailed balance for the relativistic matter. Especially, we note that the ratio can arrive at the value of the BCS theory $\omega_g/T_c\approx 3.5$ by taking proper values of $\epsilon$, $\alpha_2$, $\beta_2$ and $m$.
\end{abstract}



\maketitle

\section{Introduction}

The AdS/CFT correspondence~\cite{Maldacena,polyakov,Witten} relates
a weak coupling gravity theory in an anti-de Sitter space to a
strong coupling conformal field theory in one less dimensions.
Recently it has been applied to condensed matter physics and in
particular to superconductivity \cite{Gubser:2005ih,GubserPRD78}. In
the pioneering papers Gubser \cite{Gubser:2005ih,GubserPRD78}
suggested that near the horizon of a charged black hole there is in
operation a geometrical mechanism parameterized by a charged scalar
field of breaking a local $U(1)$ gauge symmetry. Then, the
gravitational dual of the transition from normal to superconducting
states in the boundary theory was constructed. This dual consists of
a system with a black hole and a charged scalar field, in which the
black hole admits scalar hair at temperature lower than a critical
temperature, but does not possess scalar hair at higher
temperatures~\cite{HartnollPRL101}. In this system a scalar
condensate can take place through the coupling of the scalar field
with the Maxwell field of the background. Much attention has been
focused on the application of AdS/CFT correspondence to condensed
matter physics since then
\cite{HartnollJHEP12,HorowitzPRD78,Nakano-Wen,Amado,
Koutsoumbas,Maeda79,Sonner,HartnollRev,HerzogRev,PW,
Ammon:2008fc,Gubser:2009qm,CJ0}.

Ho\v{r}ava \cite{ho1,ho2} proposed a new class of quantum gravity.
The key property of this theory is the three dimensional general
covariance and time re-parameterization invariance. It is this
anisotropic rescaling that makes Ho\v{r}ava's theory power-counting
renormalizable. Therefore, many authors pay their attention to this
gravity theory and its cosmological and astrophysical applications,
and found many interesting results
\cite{KS,CLS,CO,pia,RG,WW,CY,MK,YSM,Nis,CCO1,CJ1,CJ2,DJ0}. These
investigations imply that there exists the distinct difference
between the Ho\v{r}ava-Lifshitz theory and Einstein's gravity.

In the Ho\v{r}ava-Lifshitz gravity, Kiritsis and Kofinas \cite{Matter},  Kimpton and Padilla \cite{Matter1}  proposed the non-relativistic matter. They constructed the most general  action of matter coupled to gravity
with the foliation-preserving diffeomorphism. The action obeys the usual power-counting renormalisability conditions used in Ho\v{r}ava-Lifshitz gravity and assuming the temporal derivatives are as in the relativistic theory.

Recently, in order to see what difference will appear for the
holographic superconductivity in the Ho\v{r}ava-Lifshitz theory,
comparing with the case of the relativistic general relativity, Cai
{\it et al.} \cite{Cai-Zhang} studied the phase transition of planar
black holes in the Ho\v{r}ava-Lifshitz gravity with the condition of
the detailed balance in which the metric function is described by
$f(r)=x^2-\sqrt{c_0 x}$. They argued that the holographic
superconductivity is a robust phenomenon associated with asymptotic
AdS black holes. And they also got a relation connecting the gap
frequency in conductivity with the critical temperature, which is
given by $ \frac{\omega_g}{T_c}\approx 13, \label{Cai} $ with the
accuracy more than $93\%$ for a range of scalar masses. More recently,
Lin, Abdalla and Wang \cite{Wang}  generalized the investigation to the holographic superconductors related to the non-relativistic matter in the Schwarzschild black hole in the low energy limit of Ho\v{r}ava-Lifshitz spacetime.

Note that the Ho\v{r}ava-Lifshitz black hole without the condition of the detailed balance has rich physics \cite{Horava,LMP,Cai1}, i.e., changing the parameter of the detailed balance $\epsilon$ from
$0$ to $1$ it can produce the different  black holes for the Ho\v{r}ava-Lifshitz theory and Einstein's gravity, and the non-relativistic matter in Ho\v{r}ava-Lifshitz gravity has new properties.
In this paper we will extend the study to case of the non-relativistic matter in a Ho\v{r}ava-Lifshitz black hole without the condition of the detailed balance, and investigate how the parameter of the detailed balance and non-relativistic parameters influence on the scalar condensation formation, the electrical conductivity, and the ratio $\omega_g/T_c $ which connects the gap frequency in conductivity with the critical temperature.

The paper is organized as follows. In Sec. II we present
black hole with hyperbolic horizons in Ho\v{r}ava-Lifshitz gravity in which the action
without the condition of the detailed balance. In Sec. III we explore
the  condensation of the relativistic matter in the Ho\v{r}ava-Lifshitz black hole
background by numerical approach. In Sec. IV we
study the electrical conductivity and find ratio of the
gap frequency in conductivity to the critical temperature. We
summarize and discuss our conclusions in the last section.

\section{  black hole with hyperbolic horizon in $z=3$ Ho\v{r}ava-Lifshitz gravity}

In non-relativistic field theory, space and time have different scalings, which is called anisotropic scaling, $  x^i\rightarrow bx^i,$ $t\rightarrow b^zt,$ $ i=1,2,3,$
where $z$ is called {\it dynamical critical exponent}. In order for a theory to be power counting
renormalizable, the critical exponent has at least $z=3$ in four
spacetime dimensions.  For $z=3$, the action without the
condition of the detailed balance for the Ho\v{r}ava-Lifshitz theory
can be expressed as \cite{Horava,LMP}
\begin{eqnarray}
\label{eq6} I &=& \int dt~d^3x [{\cal L}_0 +(1-\epsilon^2){\cal
L}_1],
\end{eqnarray}
with
\begin{eqnarray}
 {\cal L}_0 &=& \sqrt{g}N \left [\frac{2}{\kappa^2}
(K_{ij}K^{ij}-\lambda K^2) +\frac{\kappa^2\mu^2 (\Lambda
R-3\Lambda^2)}{8(1-3\lambda)}\right ],  \nonumber \\ \nonumber
 {\cal L}_1  &=& \sqrt{g}N \left [\frac{\kappa^2\mu^2(1-4\lambda)}{32(1-3\lambda)}R^2
-\frac{\kappa^2}{2\omega^4}\left(C_{ij}-\frac{\mu
\omega^2}{2}R_{ij}\right)
\left(C^{ij}-\frac{\mu\omega^2}{2}R^{ij}\right) \right], \\ \nonumber
K_{ij}&=&\frac{1}{2N}(\dot g_{ij}-\nabla_iN_j-\nabla_jN_i),\\ \nonumber
C^{ij}&=&\epsilon^{ikl} \nabla _k \left (R^j_{\ l}-\frac{1}{4}R\delta^j_l\right)
= \epsilon^{ikl}\nabla_k R^j_{\ l} -\frac{1}{4}\epsilon^{ikj}\partial_kR,
\end{eqnarray}
where $\kappa^2$, $\mu$, $\Lambda$, and $\omega$  are constant
parameters,  $\epsilon$ is parameter of the detailed balance
($0<\epsilon\leq1$), $N^i$ is the shift vector, $K_{ij}$  is the
extrinsic curvature and $C_{ij}$ the Cotten tensor. It is
interesting to note that the action (\ref{eq6}) reduces to the
action in Ref. \cite{LMP} if $\epsilon=0$, and it becomes the action
for the Einstein's gravity if $\epsilon =1$.

From the action (\ref{eq6}), Cai {\it et al.}  \cite{Cai1} found a
static black hole with hyperbolic horizon whose horizon has an
arbitrary constant scalar curvature $2k$ with $\lambda=1$. The line
element of the black hole can be expressed as
\begin{equation}
\label{metric}
ds^2 = -N^2(r) dt^2 +\frac{dr^2}{f(r)} +r^2 d\Omega_k^2,
\end{equation}
with
\begin{equation}
\label{N2}
N^2 =f = k +\frac{x^2}{1-\epsilon^2}
-\frac{\sqrt{\epsilon^2 x^4+(1-\epsilon^2)c_0 x}}{1-\epsilon^2},
\end{equation}
where $x=\sqrt{-\Lambda}~r$, $k=-1,~0,~1$, and $c_0=[x_+^4+2 k
x_++(1-\varepsilon^2)k^2]/x_+$ in which $x_+$  is the horizon radius
of the black hole, i.e., the largest root of $f(r)=0$. Comparing
with the standard AdS$_4$ spacetime, we may set
$\frac{-\Lambda}{1+\epsilon}=\frac{1}{L_{AdS}^2}$, where $L_{AdS}$
is the radius of AdS$_4$. The authors in Ref. \cite{Cai1} also found
that the solution has a finite mass $M=\kappa^2 \mu^2 \Omega_k
\sqrt{-\Lambda} c_0/16$. For $\epsilon=0$, the solution goes back to
the solution in Ref. \cite{LMP}.

The Hawking temperature of the black hole is
\begin{equation}
\label{Hawking temperature}
T = \frac{\sqrt{-\Lambda}}{8\pi}
\frac{3x_+^4+2kx_+^2-(1-\epsilon^2)k^2}{x_+[x_+^2+(1-\epsilon^2)k]},
\end{equation}
which is always a monotonically increasing function of horizon
radius $x_+$ in the physical regime. This implies that the black
holes with hyperbolic horizons in the Ho\v{r}ava-Lifshitz theory are
thermodynamically stable.

\section{ Condensation for Non-relativistic matter in Ho\v{r}ava-Lifshitz gravity }

We now study the  condensation for non-relativistic matter in the Ho\v{r}ava-Lifshitz
gravity.  For the Arnowitt-Deser-Misner metric
 \bqn
 \label{metric0}
ds^2=-N^2dt^2+\gamma_{ij}\left(dx^i-N^i dt\right)\left(dx^j-N^j dt\right),
 \eqn
the Lagrangian of complex scalar and electromagnetic fields for the non-relativistic matter in the Ho\v{r}ava-Lifshitz gravity can be expressed as \cite{Matter}
 \bqn
 \label{HM1}
{\cal L}_H^E&=&\frac{2}{N^2}\gamma^{ij}\left(F_{0i}-F_{ki}N^k\right)\left(F_{0j}-F_{lj}N^l\right)
-F_{ij}F^{ij}\\ \nonumber &-&\beta_0-\beta_1a_iB^i-\beta_2B_iB^i-{\cal G}_E,\\
\label{HM11}
{\cal
L}_H^S&=&\frac{1}{2N^2}\left|\partial_t{\Psi}-N^i\partial_i\Psi\right|^2-\frac{1}{2}\left|
\partial\Psi\right|^2-\frac{1}{2}V(|{\Psi}|)+\alpha_2\left|\partial\Psi\right|^2-{\cal
H}_S,
 \eqn
with
 \bqn
 \label{HM2}
{\cal G}_E&=&\beta_3\left(B_iB^i\right)^2+\beta_4\left(B_iB^i\right)^3+\beta_5\left(\nabla_iB_j\right)\left(\nabla^iB^j\right)+\beta_6\left(B_iB^i\right)\left(\nabla_kB_j\right)\left(\nabla^kB^j\right)\nb\\
&&+\beta_7\left(\nabla_iB_j\right)\left(\nabla^iB^k\right)\left(\nabla^jB_j\right)+\beta_8\left(\nabla_i\nabla_jB_k\right)\left(\nabla^i\nabla^jB^k\right),\nb\\
{\cal
H}_S&=&\alpha_3\left(\Psi\Delta\Psi\right)^2+\alpha_4\left(\Psi\Delta\Psi\right)^3
+\alpha_5\Psi\Delta^2\Psi+\alpha_6\left(\Psi\Delta\Psi\right)\left(\Psi\Delta^2\Psi\right)
+\alpha_7\Psi\Delta^3\Psi,\nonumber \\
 \eqn
where ${\cal G}_E$ and ${\cal H}_S$ are the Ho\v{r}ava-Lifshitz higher order corrections,  $\alpha_i$ and $\beta_i$ can be taken as constants, $F_{ij}=\partial_jA_i-\partial_iA_j$,  and $B^i=\frac{1}{2} \frac{\epsilon^{ijk}}{\sqrt{\gamma}}F_{jk}$ with the Levi-Civita symbol $\epsilon^{ijk}$. In this paper, we just consider the lower order terms of above equations.

The coupling between electromagnetic field and scalar field can be constructed and then the Lagrangian $\cal L_H^S$  should be rewritten as \cite{Wang}
 \bqn
 \label{TR4}
\tilde{\cal
L}_H^S&=&\frac{1}{2N^2}\left|\partial_t{\Psi}-iqA_0\Psi-N^i\left(\partial_i\Psi-iqA_i\Psi\right)\right|^2\\  \nonumber
&-&\left(\frac{1}{2}-\alpha_2\right)\left|\partial\Psi-iqA_i\Psi\right|^2-\frac{1}{2}V(|{\Psi}|)-\tilde{\cal
H}_S,
 \eqn
where ${\cal H}_S$ is replaced by $\tilde{\cal H}_S$ with
$\partial_i\rightarrow\partial_i-iqA_i$.
Therefore, the action of coupling between complex scalar and electromagnetic fields for the non-relativistic matter in the Ho\v{r}ava-Lifshitz gravity can be taken as
 \bqn
 \label{TR5}
S_H=\int dtd^3x N\sqrt{\gamma}\left(\frac{1}{4}{\cal L}_H^E+2\tilde{\cal
L}_H^S\right),
 \eqn
which will reduce into the model in general relativity when $\alpha_i=\beta_i=0$.

In the background of the black hole described by Eq. (\ref{N2}) with $k=0$, we focus our attention on the case that these fields are weakly coupled to gravity, i.e., they do not backreact on the metric of the
spacetime. Thus, we can take the ansatz
\begin{eqnarray}
  A_{\mu}&=&(\phi(r),0,0,0),\nonumber \\ \psi&=&\psi(r).
\end{eqnarray}
This ansatz implies that the phase factor of the complex scalar
field is a constant. Therefore, we may take $\psi$ to be real. In
the background of the black hole  described by Eqs. (\ref{metric}) and
(\ref{N2}) with $k=0$, the equations of the scalar field $\psi(r)$ and the
scalar potential $\phi(r)$ are given by
\begin{eqnarray}
&&\psi^{\prime\prime}+\left(
\frac{f^\prime}{f}+\frac{2}{r}\right)\psi^\prime
+\frac{1}{1-2\alpha_2}\left(\frac{\phi^2}{f^2}-\frac{m^2}{f}\right)\psi=0, \label{Psi}
\\
&&\phi^{\prime\prime}+\frac{2}{r}\phi^\prime-\frac{2\psi^2}{f}\phi=0~,
\label{Phi}
\end{eqnarray}
where  a prime denotes the derivative  with respect to $r$.

At the event horizon $r=r_+$, we must have
\begin{eqnarray}
 \psi(r_{+})&=&-\frac{3 (1-2 \alpha_2)r_+ \psi^\prime(r_{+})}{2 m^{2} L^2},\nonumber \\  \phi(r_{+})&=&0,
\end{eqnarray}
because their norms are required to be finite, where
$L^2=L_{AdS}^2/(1+\epsilon)$ . And at the asymptotic region
($r\rightarrow\infty$), the solutions behave like
\begin{eqnarray}
\psi&=&\frac{\psi_{-}}{r^{\lambda_{-}}}+\frac{\psi_{+}}{r^{\lambda_{+}}}\,,\nonumber \\
\phi&=&\mu-\frac{\rho}{r}\,, \label{infinity}
\end{eqnarray}
with
\begin{eqnarray}
\lambda_\pm=\frac{1}{2}\left(3\pm\sqrt{9+\frac{4m^{2}L_{AdS}^2}{1-2 \alpha_2}}~\right)\,,\label{LambdaZF}
\end{eqnarray}
where $\mu$ and $\rho$ are interpreted as the chemical potential and
charge density in the dual field theory, respectively. Because the
boundary is a (2+1)-dimensional field theory, $\mu$ is of mass
dimension one and $\rho$ is of mass dimension two. We can read off
the expectation values of operator $\mathcal{O}$ dual to the field
$\psi$. From Ref. \cite{kw}, we know that for $\psi$, both of these
falloffs are normalizable, and in order to keep the theory stable,
we should either impose
 \begin{eqnarray} \psi{_{-}}=0,\quad \text{and} \quad \langle
 \mathcal{O}_{+}\rangle=\psi{_{+}},\end{eqnarray}
or
 \begin{eqnarray} \psi{_{+}}=0,\quad \text{and} \quad \langle
 \mathcal{O}_{_{-}}\rangle=\psi{_{-}}.\end{eqnarray}
Note that the dimension of temperature $T$ is of mass dimension one,
the ratio $T^2/\rho$ is dimensionless. Therefore increasing $\rho$
while $T$  is fixed, is equivalent to decrease $T$ while $\rho$ is
fixed. In our calculation, we find that when $\rho>\rho_c$, the
operator condensate will appear; this means when $T<T_c$ there will
be an operator condensate, that is to say, the superconducting phase
occurs. We will impose boundary condition $\psi{_{-}}=0$ in the following discussion.

\begin{figure}
\includegraphics[scale=1.18]{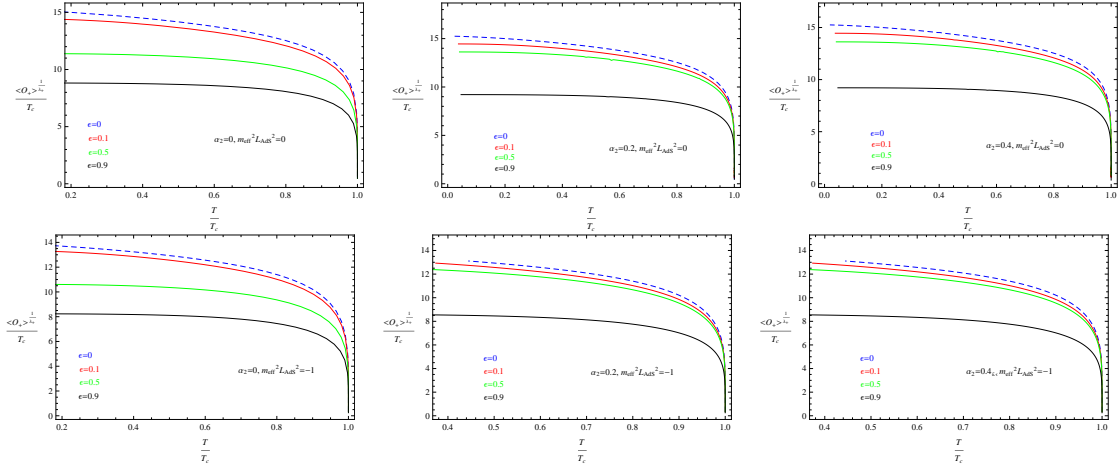}\hspace{0.2cm}%
\caption{\label{condensate} (Color online) The condensate as a
function of the temperature with fixed values
$m_{eff}^{2}L_{AdS}^{2}=0,~-1$ and arbitrary $\beta_2$. The four lines from top to bottom
correspond to increasing $\epsilon$, i.e., $0$ (blue), $0.1$ (red),
$0.5$ (green) and $0.9$ (black), respectively. It is shown that the
condensation gap becomes smaller as $\epsilon$ (or $\alpha_2$) increases for the
same $m_{eff}^{2}$.}
\end{figure}

\begin{figure}
\includegraphics[scale=1.25]{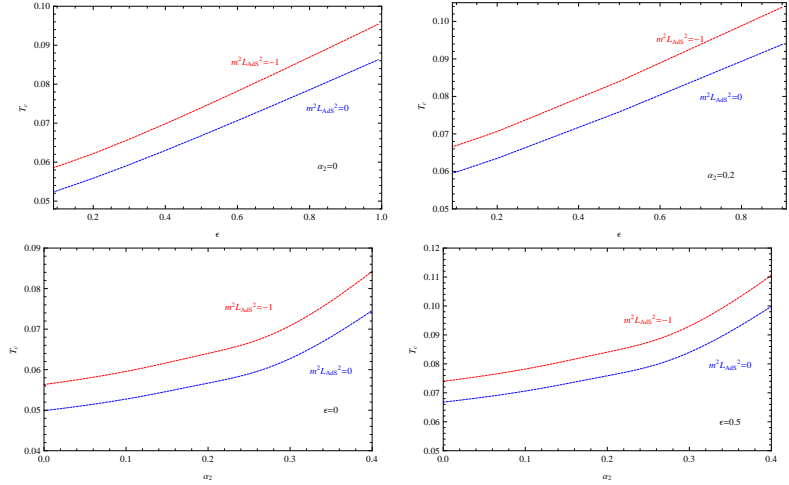}\vspace{0.0cm}
\caption{\label{HLTc} (Color online) The critical temperature as a
function of the balance parameter (or non-relativistic parameter) with fixed values
$m_{eff}^{2}L_{AdS}^{2}$ and arbitrary $\beta_2$. The two lines from top to bottom correspond to
$m_{eff}^{2}L_{AdS}^{2}=-1$ (red) and $0$ (blue), respectively.}
\end{figure}

Eqs. (\ref{Psi}) and (\ref{Phi}) can be solved numerically
by doing integration from the horizon out to the infinity with the
boundary  conditions mentioned above.
Changing the values of the balance parameter $\epsilon$ and non-relativistic parameter $\alpha_2$, we present in Fig. \ref{condensate} the influence of the parameters $\epsilon$ and $\alpha_2$ on the condensation with fixed values $ m_{eff}^2 L_{AdS}^2=0$, $-1$ (here and hereafter $m_{eff}^2=\frac{m^2}{1-2\alpha_2}$) and arbitrary $\beta_2$,  and in Fig. \ref{HLTc} the critical temperature  as a function of the balance parameter and non-relativistic parameter $\alpha_2$ with fixed values  $ m_{eff}^2 L_{AdS}^2=0$, $-1$ and arbitrary $\beta_2$. In table \ref{Tc-D6} we present the critical temperature obtained by the numerical method.
We know from the figures and the table that as the parameter of the
detailed balance increases with fixed non-relativistic parameter $\alpha_2$ and effective mass of the scalar field, the condensation gap becomes smaller, corresponding to larger the critical temperature, which means that the scalar hair can be formed
easier for the larger $\epsilon$. Similarly, the scalar hair can be formed easier as the non-relativistic parameter $\alpha_2$ becomes larger with fixed balance parameter and effective mass of the scalar field.
And the figures and table also show that, for the same $\epsilon $ or $\alpha_2$, the condensation gap
becomes larger if $m_{eff}^2$ becomes less negative, which means that it
is harder for the scalar hair to form as the effective mass of the scalar field becomes larger. We should point out that the parameter $\beta_2$ dose not affect the condensation in this model.

\begin{table}[htbp]
\begin{center}
\caption{\label{Tc-D6} The critical temperature $T_{c}$  obtained by numerical method.}
\begin{tabular}{c | c | c | c | c | c}
         \hline
\multicolumn{6}{c}{$\alpha_2=0$}\\
    \hline
 ~~&$\epsilon=0.0$ ~~&$\epsilon=0.1$~~&$\epsilon=0.2$~~& $\epsilon=0.5$~~& $\epsilon=0.9$
        \\
        \hline
$m_{eff}^2L_{AdS}^2=0~$~~~&~~$0.0499$~~&~~$0.0502$~~&~~$0.0510$~~&~~$0.0545$~~&~~$0.0599$
          \\
$m_{eff}^2L_{AdS}^2=-1$~~&~~$0.0563$~~&~~$0.0589$~~&~
 ~$0.0622$~~&~~$0.0740$~~&~~$0.0914$
          \\
    \hline

          \hline
\multicolumn{6}{c}{$\alpha_2=0.1$}\\
\hline

$m_{eff}^2L_{AdS}^2=0$~~&~~$0.0527$~~&~~$0.0557$~~&~~
 $0.0591$~~&~~$0.0706$~~&~~$0.0874$~~
          \\
 $m_{eff}^2L_{AdS}^2=-1$~~&~~$0.0596$~~&~~$0.0623$~~&~~
 $0.0657$~~&~~$0.0782$~~&~~$0.0966$~~
          \\
\hline

   \hline
\multicolumn{6}{c}{$\alpha_2=0.2$}\\
\hline

$m_{eff}^2L_{AdS}^2=0$~~&~~$0.0567$~~&~~$0.0598$~~&~~
 $0.0635$~~&~~$0.0759$~~&~~$0.0938$~~
          \\
 $m_{eff}^2L_{AdS}^2=-1$~~&~~$0.0640$~~&~~$0.0669$~~&~~
 $0.0707$~~&~~$0.0840$~~&~~$0.1038$~~
          \\
\hline

 \hline
\multicolumn{6}{c}{$\alpha_2=0.4$}\\
\hline

$m_{eff}^2L_{AdS}^2=0$~~&~~$0.0746$~~&~~$0.0788$~~&~~
 $0.0835$~~&~~$0.0999$~~&~~$0.1235$~~
          \\
 $m_{eff}^2L_{AdS}^2=-1$~~&~~$0.0842$~~&~~$0.0881$~~&~~
 $0.0930$~~&~~$0.1106$~~&~~$0.1366$~~
          \\
\hline
\end{tabular}
\end{center}
\end{table}

\section{Electrical Conductivity in Ho\v{r}ava-Lifshitz black-hole background }

In the study of (2+1) and (3+1)-dimensional superconductors in Einstein gravity,
Horowitz {\it et al.} \cite{HorowitzPRD78} got a universal relation
connecting the gap frequency in conductivity with the critical
temperature $T_c$, which is described by
\begin{eqnarray}
\frac{\omega_g}{T_c}\approx 8,
\end{eqnarray}
with deviations of less than $8\%$. This is roughly twice the BCS
value 3.5 indicating that the holographic superconductors are
strongly coupled. The authors in Refs. \cite{PW,Gregory}
found that this relation is not stable in the presence of the
Gauss-Bonnet correction terms. And Cai {\it et al.} \cite{Cai-Zhang}
got a relation
\begin{eqnarray}
\frac{\omega_g}{T_c} \approx 13, \end{eqnarray}
with the accuracy more than $93\%$ for a planar Ho\v{r}ava-Lifshitz black hole  with the condition of the
detailed balance for the relativistic matter.

We now study this relation for the  non-relativistic matter in the Ho\v{r}ava-Lifshitz gravity.
In order to compute the electrical conductivity, we should study the electromagnetic perturbation in this Ho\v{r}ava-Lifshitz black hole background, and then calculate the linear response to the
perturbation. In the probe approximation, the effect of the
perturbation of metric can be ignored. Assuming that the
perturbation of the vector potential is translational symmetric and
has a time dependence as $\delta A_x=A_x(r)e^{-i\omega  t}$, we find
that the equation of  motion for $A_x$ in the Ho\v{r}ava-Lifshitz black hole
background reads
\begin{eqnarray}
A_{x}^{\prime\prime}+\frac{f^\prime}{f}A_{x}^\prime
+\frac{2}{2+\beta_2}\left[\frac{\omega^2}{f^2}-\frac{2(1-2\alpha_2)\psi^2}{f}\right]A_{x}=0 \; ,
\label{Maxwell Equation}
\end{eqnarray}
where  a prime denotes the derivative  with respect to $r$. An
ingoing wave  boundary condition near the horizon is given by
\begin{eqnarray}
A_{x}(r)\sim f(r)^{-\frac{2 i \omega L^2}{3 r_+ \sqrt{1+\beta_2/2}}}. \end{eqnarray} In
the asymptotic AdS region ($r\rightarrow\infty$),  the general
behavior should be
\begin{eqnarray}\label{Maxwell boundary}
A_{x}=A^{(0)}+\frac{A^{(1)}}{r} +\cdots~.
\end{eqnarray}
By using AdS/CFT correspondence and the Ohm's law, we know that the
conductivity can be expressed as
  \cite{HorowitzPRD78}
\begin{eqnarray}\label{GBConductivity}
\sigma=\frac{\langle J_x \rangle}{E_x}=-\frac{i\langle J_x
\rangle}{\omega A_x}=\frac{A^{(1)}}{i\omega A^{(0)}} \, .
\end{eqnarray}
\begin{figure}
\includegraphics[scale=1.1]{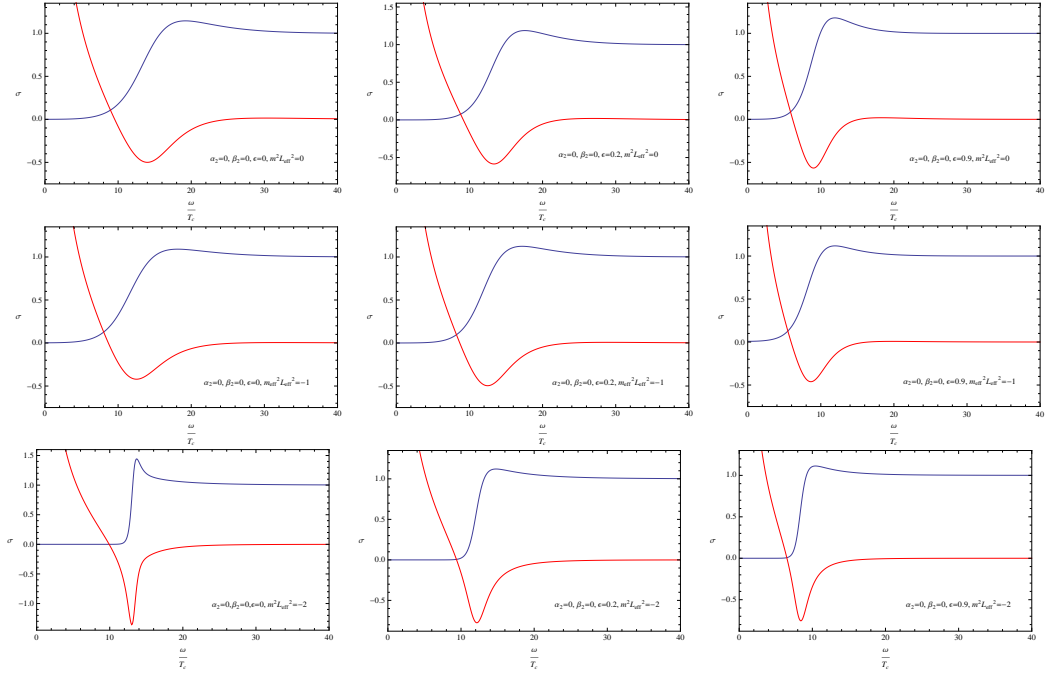}\hspace{0.2cm}%
\caption{\label{Conductivity} (Color online) The conductivity of the
superconductors for $\epsilon=0,~0.2$ and $0.9$ with  $\beta_2=0$,
$\alpha_2=0$ and $m_{eff}^{2}L_{AdS}^{2}=0,~-1,~-2$. The solid (blue) line represents
the real part of the conductivity, $Re(\sigma)$, and dashed (red)
line is the imaginary part of the conductivity, $Im(\sigma)$.}
\end{figure}

\begin{figure}
\includegraphics[scale=1.1]{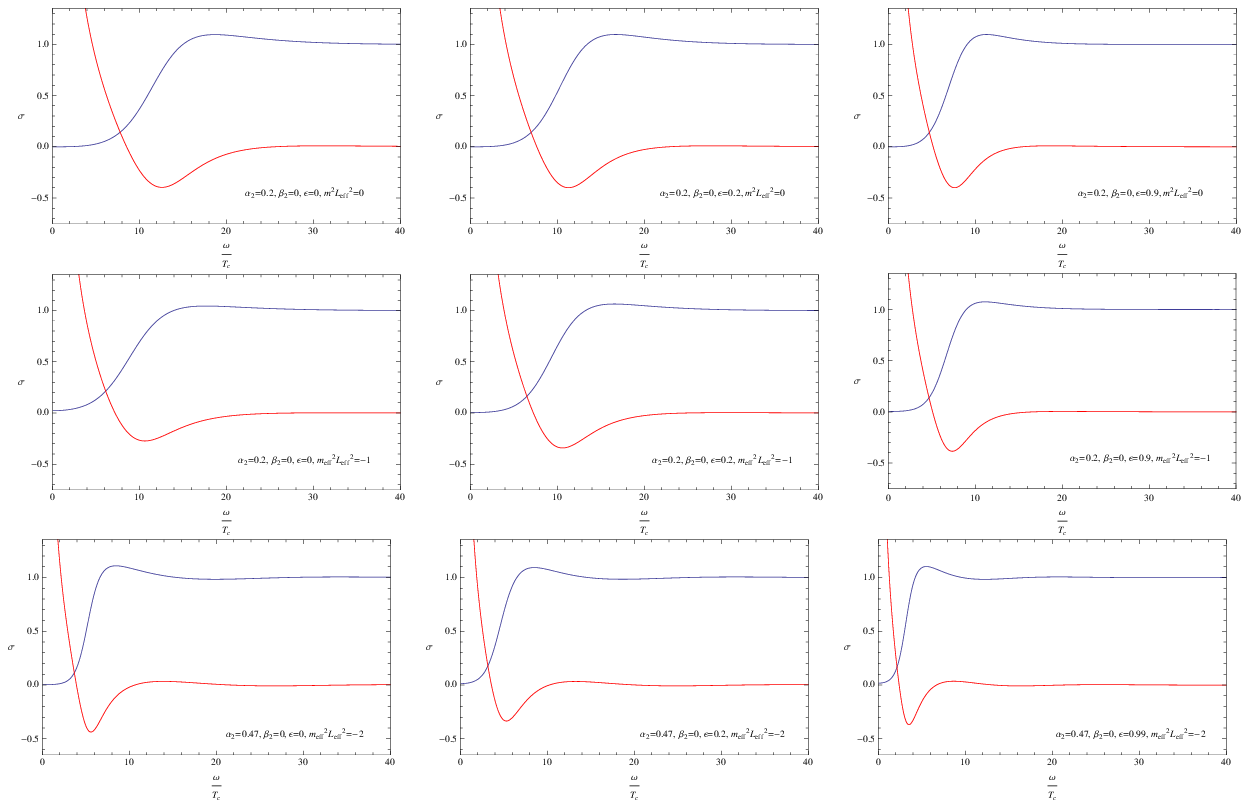}\hspace{0.2cm}%
\caption{\label{Conductivity1} (Color online) The conductivity of the
superconductors for $\epsilon=0,~0.2$, $0.9$ and $0.99$ with $\beta_2=0$,
$\alpha_2=0.2,~0.47$ and $m_{eff}^{2}L_{AdS}^{2}=0,~-1~,-2$. The solid (blue) line represents
the real part of the conductivity, $Re(\sigma)$, and dashed (red)
line is the imaginary part of the conductivity, $Im(\sigma)$.}
\end{figure}

\begin{figure}
\includegraphics[scale=1.1]{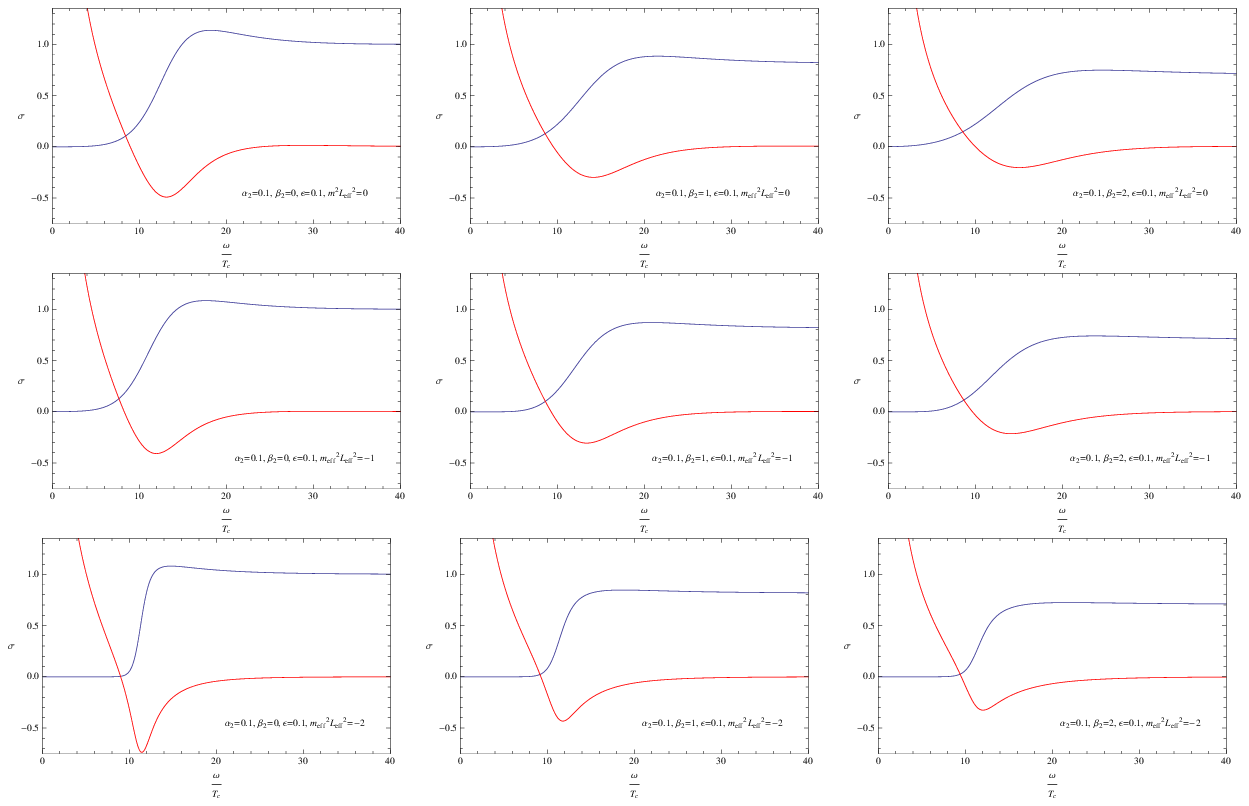}\hspace{0.2cm}%
\caption{\label{Conductivity2} (Color online) The conductivity of the
superconductors for $\beta_2=0,~1$ and $2$ with
$\alpha_2=0.1$, $\epsilon=0.1$ and $m_{eff}^{2}L_{AdS}^{2}=0,~-1~,-2$. The solid (blue) line represents
the real part of the conductivity, $Re(\sigma)$, and dashed (red)
line is the imaginary part of the conductivity, $Im(\sigma)$.}
\end{figure}

%\begin{table}[htbp]
\begin{table}
\caption{\label{ConductivityTc} The ratio $\omega_{g}/T_{c}$ for
different values of the parameters $\epsilon$, $\alpha_2$ and $\beta_2$ with
$m_{eff}^{2}L_{AdS}^{2}=0,~-1$ and $-2$. }
\begin{tabular}{|c|c|c|c|c|c|c|}
\hline
\multicolumn{7}{|c|}{$\alpha_2=0$,~~~$\beta_2=0$}\\
         \hline
$~~~~$ &$\epsilon=0$&$\epsilon=0.1$&$\epsilon=0.2$&$\epsilon=0.5$&$\epsilon=0.9$&$\epsilon=0.99$
          \\
        \hline
$m_{eff}^{2}L_{AdS}^{2}=0$ & ~~$14.6$~~ & ~~$13.9$~~ & ~~$13.2$~~&
~~$11.3$~~& ~~$9.0$~~& ~~$8.6$~~
          \\
        \hline
$m_{eff}^{2}L_{AdS}^{2}=-1$ & $13.8$ & $13.2$ & $12.6$ & $10.8$ & $8.6$ &
$8.2$
          \\
        \hline
$m_{eff}^{2}L_{AdS}^{2}=-2$ & $12.9$ & $12.4$ & $11.9$ & $10.3$ & $8.4$ &
$8.1$
          \\
         \hline
\hline
\multicolumn{7}{|c|}{$\alpha_2=0.2$,~~~$\beta_2=0$}\\
         \hline
$~~~$ &$\epsilon=0$&$\epsilon=0.1$&$\epsilon=0.2$&$\epsilon=0.5$&$\epsilon=0.9$&$\epsilon=0.99$
          \\
        \hline
$m_{eff}^{2}L_{AdS}^{2}=0$ & ~~$13.0$~~ & ~~$11.8$~~ & ~~$11.5$~~&
~~$~9.7~$~~& ~~$~7.8~$~~& ~~$~7.2~$~~
          \\
\hline
$m_{eff}^{2}L_{AdS}^{2}=-1$ & $11.2$ & $10.5$ & $10.1$ & $~9.1~$ & $7.2$ &
$~6.9~$
         \\
\hline
$m_{eff}^{2}L_{AdS}^{2}=-2$ & $ 10.3$ & $ 9.7$ & $9.3 $ & $~ 8.6~$ & $ 6.8 $ &
$~6.5~$\\
        \hline
\hline
\multicolumn{7}{|c|}{$\alpha_2=0.47$,~~~$\beta_2=0$}\\
         \hline
$~~~$ &$\epsilon=0$&$\epsilon=0.1$&$\epsilon=0.2$&$\epsilon=0.5$&$\epsilon=0.9$&$\epsilon=0.99$
          \\
        \hline
$m_{eff}^{2}L_{AdS}^{2}=0$ & ~~$ 6.9 $~~ & ~~$ 6.4 $~~ & ~~$ 5.8 $~~&
~~$~ 5.0 ~$~~& ~~$~ 4.3 ~$~~& ~~$~ 4.1 ~$~~
  \\
        \hline
$m_{eff}^{2}L_{AdS}^{2}=-1$ & ~~$  6.5$~~ & ~~$ 5.9 $~~ & ~~$ 5.4 $~~&
~~$~ 4.8 ~$~~& ~~$~ 4.0 ~$~~& ~~$~ 3.8 ~$~~
  \\
        \hline
$m_{eff}^{2}L_{AdS}^{2}=-2$ & ~~$5.6$~~ & ~~$5.3$~~ & ~~$5.0$~~&
~~$~4.6~$~~& ~~$~3.7~$~~& ~~$~3.5~$~~      \\  \hline
\hline
\multicolumn{7}{|c|}{$\alpha_2=0.1$,~~~$\epsilon=0.1$}\\
         \hline
$~~~~$ &$\beta_2=0$&$\beta_2=0.5$&$\beta_2=1$&$\beta_2=1.5$&$\beta_2=2$&$\beta_2=2.5$
          \\
        \hline
$m_{eff}^{2}L_{AdS}^{2}=0$ & ~~$13.0$~~ & ~~$13.5$~~ & ~~$14.0$~~&
~~$14.5$~~& ~~$15.0$~~& ~~$~15.8~$~~
          \\
        \hline
$m_{eff}^{2}L_{AdS}^{2}=-1$ & $12.2$ & $12.8$ & $13.2$ & $13.6$ & $14.0$ &
$~14.5~$
          \\
        \hline
$m_{eff}^{2}L_{AdS}^{2}=-2$ & $11.4$ & $11.8$ & $12.1$ & $12.3$ & $12.5$ &
$~12.8~$
          \\
         \hline

\end{tabular}
\end{table}

In Figs. \ref{Conductivity} and \ref{Conductivity1} we plot the frequency dependent
conductivity obtained by solving the equation of motion (\ref{Maxwell Equation}) numerically
for $\epsilon=0,~0.5$ and $0.9$ (or $0.99$) with  $\beta_2=0$,  $\alpha_2=0,~0.2,~0.47$ and $m_{eff}^{2}L_{AdS}^{2}=0,~-1$
and $-2$.  We find that, for the same value of $m_{eff}^2 L_{AdS}^2$, the gap
frequency $\omega_{g}$ decreases with the increase of the parameters
$\epsilon$ or $\alpha_2$. In each plot, the real part of the conductivity,
Re[$\sigma$], approaches to a limit when the frequency grows. The
limit for the case $\epsilon=0$ and $\alpha_2=0$ is one, but generally it increases as
parameters $\epsilon$ or $\alpha_2$ increases.  The imaginary part of conductivity
Im[$\sigma$] becomes zero when $\omega\rightarrow \infty$, but it
goes to infinity when the frequency approaches zero.

In Fig. \ref{Conductivity2} we plot the frequency dependent conductivity  for $\beta_2=0,~1$ and $2$ with
$\alpha_2=0.1$, $\epsilon=0.1$ and $m_{eff}^{2}L_{AdS}^{2}=0,~-1~,-2$.
We note that, for the same values of $\epsilon$, $\alpha_2$ and $m_{eff}^2 L_{AdS}^2$, the gap
frequency $\omega_{g}$ increases with the increase of the parameters $\beta_2$. That is to say, the ratio of the gap frequency in conductivity $\omega_g$ to the critical temperature $T_c$ increases as the parameters $\beta_2$ increases with fixed $\alpha_2$, $\epsilon$ and $m_{eff}^2$.

In table \ref{ConductivityTc} we also present how the ratio $\omega_{g}/T_{c}$  relate to the balance parameter and non-relativistic parameter with fixed values $m_{eff}^2 L_{AdS}^{2}=0$, $-1$ and $-2$, which shows that the ratio $\omega_{g}/T_c$ decreases with the increase of the balance parameter or the non-relativistic parameter $\alpha_2$, but increases with the increase of the parameter $\beta_2$.

From Figs. \ref{Conductivity},  \ref{Conductivity1} and \ref{Conductivity2} and table \ref{ConductivityTc}, we find that the ratio of the gap frequency in conductivity $\omega_g$ to the critical temperature $T_c$ in this black hole reduces to Cai's result $\omega_g/T_c\approx 13$ \cite{Cai-Zhang} found in the Ho\v{r}ava-Lifshitz gravity with the  condition of the detailed balance for the relativistic matter when  $\epsilon=0$, $\beta_2=0$ and $\alpha_2=0$, while it tends to the Horowitz-Roberts relation $\omega_g/T_c\approx 8$ obtained in the Einstein gravity as $\epsilon \rightarrow 1$ with $\alpha_2=0$ and $\beta_2=0$.  Especially, the ratio can arrive at the value of the BCS theory $\omega_g/T_c\approx 3.5$ if we take right value for $\epsilon$, $\alpha_2$, $\beta_2$ and $m_{eff}^2$, say $\epsilon=0.99$, $\alpha_2=0.47$, $\beta_2=0$ and $m_{eff}^2 L_{AdS}^{2}=-2$.

\section{conclusions}

The behavior of the holographic superconductors  in the Ho\v{r}ava-Lifshitz gravity has been investigated in this manuscript by introducing the non-relativistic scalar and electromagnetic fields in a
planar black-hole background. We first present a detailed analysis
of the condensation of the operator $\mathcal {O}_{+}$ by the
numerical  method for the Ho\v{r}ava-Lifshitz black
hole without the condition of the detailed balance. It is found that, as the
parameter of the detailed balance $\epsilon$ increases with fixed the non-relativistic parameter $\alpha_2$ and effective  mass of the scalar field $m_{eff}^2$, the condensation gap becomes smaller,
corresponding to the larger critical temperature, which means that
the scalar hair can be formed easier for the larger $\epsilon$.
Similarly, the scalar hair can be formed easier as the non-relativistic parameter $\alpha_2$ becomes larger with fixed  detailed balance $\epsilon$ and  effective mass of the scalar field.
And it is also shown that, for the same $\epsilon $ or $\alpha_2$, the condensation gap
becomes larger if $m_{eff}^2$ becomes less negative, which means that it
is harder for the scalar hair to form as the effective mass of the scalar
field becomes larger. It is interesting to note that the parameter $\beta_2$ does not affect the condensation. We then studied the electrical conductivity for the non-relativistic matter in
the Ho\v{r}ava-Lifshitz black-hole background and find that the ratio of the gap frequency in conductivity to the critical temperature, $\omega_{g}/T_c$,  decreases with the increase of the balance parameter $\epsilon$ or the non-relativistic parameter $\alpha_2$, but increases with the increase of the parameter $\beta_2$.
The ratio reduces to Cai's result $\omega_g/T_c\approx 13$ \cite{Cai-Zhang}
found in a Ho\v{r}ava-Lifshitz gravity with the condition of the detailed balance for the relativistic matter when $\epsilon= 0$,  $\alpha_2=0$ and $\beta_2=0$, while it tends to the Horowitz-Roberts relation $\omega_g/T_c\approx 8$ \cite{HorowitzPRD78}  obtained in the Einstein gravity if we take $\epsilon \rightarrow 1$, $\alpha_2=0$ and  $\beta_2=0$.  Especially, the ratio can arrive at the value of the BCS theory $\omega_g/T_c\approx 3.5$ if we take right values of $\epsilon$, $\alpha_2$, $\beta_2$ and $m$.

\begin{acknowledgments}

This work is supported by the  National Natural Science Foundation
of China under Grant Nos. 11175065, 11475061;  the SRFDP under
Grant No. 20114306110003. S. Chen's work was partially supported by the National Natural Science Foundation of China under Grant No. 11275065. And Q. Pan's work was partially supported by the National Natural Science Foundation of China under Grant No. 11275066.
\end{acknowledgments}


\begin{thebibliography}{99}


\bibitem{Maldacena}
J. Maldacena,
%The Large N Limit of Superconformal Field Theories and Supergravity,
 Adv. Theor. Math. Phys. {\bf 2}, 231 (1998).

\bibitem{polyakov} S. S. Gubser, I. R. Klebanov and A. M. Polyakov,
%Gauge Theory Correlators from Noncritical String Theory,
Phys. Lett.
{\bf B 428}, 105 (1998). [hep-th/9802109]

\bibitem{Witten}
E. Witten,
%Anti De Sitter Space And Holography,
Adv. Theor. Math.
Phys. {\bf 2}, 253 (1998).


%\cite{Gubser:2005ih}
\bibitem{Gubser:2005ih}
  S.~S.~Gubser,
 %Phase transitions near black hole horizons,
  Class.\ Quant.\ Grav.\  {\bf 22}, 5121 (2005).
  %[arXiv:hep-th/0505189].
  %%CITATION = CQGRD,22,5121;%%


  \bibitem{GubserPRD78}
S. S. Gubser,
%Breaking an Abelian gauge symmetry near a black hole horizon,
Phys. Rev. D {\bf 78}, 065034 (2008).

\bibitem{HartnollPRL101}
S. A. Hartnoll, C. P. Herzog, and G. T. Horowitz,
%Building an AdS/CFT superconductor,
Phys. Rev. Lett. {\bf 101}, 031601 (2008).


\bibitem{HartnollJHEP12}
S. A. Hartnoll, C. P. Herzog, and G. T. Horowitz,
%Families of IIB duals for nonrelativistic CFTs,
 J. High Energy
Phys. {\bf 0812}, 015 (2008).



\bibitem{HorowitzPRD78}
G. T. Horowitz and M. M. Roberts,
%Holographic Superconductors with Various Condensates,
Phys. Rev. D {\bf 78}, 126008 (2008).

\bibitem{Nakano-Wen}
E. Nakano and Wen-Yu Wen,
%Critical magnetic field in AdS/CFT superconductor,
Phys. Rev. D {\bf 78}, 046004 (2008).

\bibitem{Amado}
I. Amado, M. Kaminski, and K. Landsteiner,
%Hydrodynamics of Holographic Superconductors,
J. High Energy Phys. {\bf 0905}, 021
(2009).

\bibitem{Koutsoumbas}
G. Koutsoumbas, E. Papantonopoulos and G. Siopsis,
%Exact Gravity Dual of a Gapless Superconductor,
J. High Energy Phys. {\bf 0907},
026 (2009).

\bibitem{Maeda79}
K. Maeda, M. Natsuume, and T. Okamura,
%Universality class of holographic superconductors,
Phys. Rev. D {\bf 79}, 126004 (2009).

\bibitem{Sonner}
Julian Sonner,
%A Rotating Holographic Superconductor,
Phys. Rev. D
{\bf 80}, 084031 (2009).

\bibitem{HartnollRev}
S. A. Hartnoll,
%Lectures on holographic methods for condensed matter physics,
Class. Quant. Grav.{\bf 26}, 224002 (2009) [arXiv: 0903.3246.]

\bibitem{HerzogRev}
C. P. Herzog,
%Lectures on Holographic Superfluidity and Superconductivity,
J. Phys. A {\bf 42}, 343001 (2009).

\bibitem{PW}
Qiyuan Pan, Bin Wang, Eleftherios Papantonopoulos, J. Oliveira, and
A. Pavan,
% Holographic Superconductors with various condensates in Einstein-Gauss-Bonnet gravity,
Phys. Rev. D{\bf 81}, 106007 (2010). arXiv: 0912.2475.
%\cite{Ammon:2008fc}

\bibitem{Ammon:2008fc}
  M.~Ammon, J.~Erdmenger, M.~Kaminski, and P.~Kerner,
%  Superconductivity from gauge/gravity duality with flavor,
  Phys.\ Lett.\  B {\bf 680}, 516 (2009).
  %[arXiv:0810.2316 [hep-th]].
  %%CITATION = PHLTA,B680,516;%%

%\cite{Gubser:2009qm}
\bibitem{Gubser:2009qm}
  S.~S.~Gubser, C.~P.~Herzog, S.~S.~Pufu, and T.~Tesileanu,
% Superconductors from Superstrings,
  Phys.\ Rev.\ Lett.\  {\bf 103}, 141601 (2009).
  %[arXiv:0907.3510 [hep-th]].
  %%CITATION = PRLTA,103,141601;%%


\bibitem{CJ0}
Songbai Chen, Liancheng Wang, Chikun Ding, and Jiliang Jing,
% Holographic superconductors in the AdS black hole spacetime with a global monopole,
Nucl. Phys. B{\bf 836}, 222 (2010). [arXiv: 0912.2397]

\bibitem{ho1}
P.~Horava,
% Quantum Gravity at a Lifshitz Point,
  Phys.\ Rev.\  D {\bf79}, 084008 (2009). [arXiv: 0901.3775]
  %%CITATION = PHRVA,D79,084008;%%

\bibitem{ho2}
  P.~Horava,
% Membranes at Quantum Criticality,
  JHEP {\bf0903}, 020 (2009). [arXiv: 0812.4287]
  %%CITATION = JHEPA,0903,020;%%



\bibitem{KS}
  A.~Kehagias and K.~Sfetsos,
%The black hole and FRW geometries of non-relativistic gravity,
   Phys. Lett. B {\bf678}, 123 (2009). [arXiv: 0905.0477]
  %%CITATION = ARXIV:0905.0477;%%

\bibitem{CLS}
  R.~G.~Cai, Y.~Liu, and Y.~W.~Sun,
% On the z=4 Horava-Lifshitz Gravity,
  JHEP {\bf0906}, 010 (2009). [arXiv: 0904.4104]
  %%CITATION = ARXIV:0904.4104;%%

\bibitem{CO}
  R.~G.~Cai and N.~Ohta,
% Horizon Thermodynamics and Gravitational Field  Equations in Horava-Lifshitz Gravity ,
 Phys. Rev. D {\bf 81}, 084061 (2010).  [arXiv:0910.2307]
  %%CITATION = ARXIV:0910.2307;%%


\bibitem{pia}
Y.~S.~Piao,
%Primordial Perturbation in Horava-Lifshitz Cosmology,
   Phys. Lett. B {\bf681}, 1 (2009). [arXiv: 0904.4117]
  %%CITATION = ARXIV:0904.4117;%%

\bibitem{RG}  Remo Garattini,
%The cosmological constant as an eigenvalue of the Hamiltonian constraint in Horava-Lifshits
%theory,
arXiv: 0912.0136.

\bibitem{WW}
  Anzhong Wang and Yumei Wu,
% Thermodynamics and classification of cosmological  models in the Horava-Lifshitz theory of gravity,
  JCAP {\bf0907}, 012 (2009). [arXiv: 0905.4117]
  %%CITATION = ARXIV:0905.4117;%%

\bibitem{CY}
  E.~O.~Colgain and H.~Yavartanoo,
% Dyonic solution of Horava-Lifshitz Gravity,
  JHEP {\bf0908}, 021 (2009). [arXiv: 0904.4357]
  %%CITATION = ARXIV:0904.4357;%%

\bibitem{MK}
  Y.~S.~Myung and Y.~W.~Kim,
% Thermodynamics of Ho\v{r}ava-Lifshitz black holes,
  arXiv: 0905.0179.
  %%CITATION = ARXIV:0905.0179;%%

\bibitem{YSM}
  Y.~S.~Myung,
% Entropy of black holes in the deformed Horava-Lifshitz gravity,
Phys. Lett. B{\bf 684}, 158 (2010).  [arXiv: 0908.4132]
  %%CITATION = ARXIV:0908.4132v2;%%

\bibitem{Nis}
  T.~Nishioka,
% Horava-Lifshitz Holography,
  Class. Quant. Grav. {\bf26}, 242001 (2009)
  [arXiv: 0905.0473].
  %%CITATION = ARXIV:0905.0473;%%

\bibitem{CCO1}
  R.~G.~Cai, L.~M.~Cao, and N.~Ohta,
% Thermodynamics of Black Holes in Horava-Lifshitz Gravity,
  Phys. Lett. B {\bf679}, 504 (2009). [arXiv: 0905.0751]
  %%CITATION = ARXIV:0905.0751;%%

\bibitem{CJ1}
  Songbai Chen and Jiliang Jing,
% Quasinormal modes of a black hole in the deformed H¨¯rava-Lifshitz gravity ,
Phys. Lett. B {\bf 687}, 124 (2010).  [arXiv: 0905.1409]
  %%CITATION = ARXIV:0905.1409;%%

\bibitem{CJ2}
  Songbai Chen and Jiliang Jing,
% Strong field gravitational lensing in the deformed H¨¯rava-Lifshitz black hole ,
  Phys. Rev. D {\bf80}, 024036 (2009). [arXiv: 0905.2055]
  %%CITATION = ARXIV:0905.2055;%%

\bibitem{DJ0}
  Chikun Ding, Songbai Chen, and Jiliang Jing,
% Dynamical evolution of scalar perturbation in Hoava-Lifshitz black-hole spacetimes,
   Phys. Rev.  D81, 024028 (2010).
  %%CITATION = ARXIV:0905.2055;%%



\bibitem{Matter} E. Kiritsis and G. Kofinas,
%Horava-Lifshitz Cosmology,
Nucl.Phys.B821 467 (2009).


\bibitem{Matter1}
 I. Kimpton and A. Padilla,
% Matter in Ho\v{r}ava-Lifshitz gravity,
 J. High Energy Phys. 04, 133 (2013)
arXiv:1301.6950.

\bibitem{Cai-Zhang}
Rong-Gen Cai and Hai-Qing Zhang,
% Holographic Superconductors with Hoava-Lifshitz Black Holes,
Phys. Rev. D81, 066003 (2010).


\bibitem{Wang} K. Lin, E. Abdalla, A. Z. Wang,
% Holographic superconductors in Ho¡¦rava-Lifshitz gravity,
[arXiv: 1406.4721]


\bibitem{Horava}  P. Horava,
% Quantum Gravity at a Lifshitz Point,
Phys. Rev. D 79, 084008 (2009).

\bibitem{LMP}  H.~Lu, J.~Mei, and C.~N.~Pope,
% Solutions to Horava Gravity,
Phys. Rev. Lett. {\bf103}, 091301 (2009).  [arXiv: 0904.1595]
  %%CITATION = ARXIV:0904.1595;%%

\bibitem{Cai1}  Rong-Gen Cai, Li-Ming Cao, and Nobuyoshi Ohta,
% Topological Black Holes in Horava-Lifshitz Gravity,
Phys. Rev. D {\bf80}, 024003 (2009). [arXiv: 0904.3670]




\bibitem{kw} I.R. Klebanov and E. Witten,
% AdS/CFT Correspondence and Symmetry Breaking,
Nucl. Phys. B {\bf
556}, 89 (1999). [hep-th/9905104]



\bibitem{Gregory}
R. Gregory, S. Kanno, and J. Soda,
% Holographic superconductors with higher curvature corrections,
J. High Energy Phys. {\bf 0910}, 010
(2009).




\end{thebibliography}
\end{document}